\documentclass[varenna]{cimento}

\usepackage{amsmath}
\usepackage{amsfonts}

\usepackage{bm}
\renewcommand{\vec}{\bm}

\newcommand{\dif}{\mathrm{d}}
\newcommand{\mi}{\mathrm{i}}
\newcommand{\me}{\mathrm{e}}

\usepackage{graphicx}

\title{Two-component ultracold Bose gases with spin-orbit coupling}

\author{G.~I. Martone}
\institute{CNR NANOTEC, Institute of Nanotechnology, Via Monteroni, 73100 Lecce, Italy}
\institute{INFN, Sezione di Lecce, 73100 Lecce, Italy}

\author{S.~Stringari}
\institute{Pitaevskii BEC Center, CNR-INO and Dipartimento di Fisica, Universit\`a di Trento, 38123 Trento, Italy}
\institute{Trento Institute for Fundamental Physics and Applications, INFN, 38123 Trento, Italy}

\shortauthor{G.~I. Martone \atque S.~Stringari}

\begin{document}

\maketitle

\begin{abstract}
These lecture notes provide an introduction to Bose-Einstein condensates with Raman-induced spin-orbit coupling. Owing
to the interplay between the peculiar single-particle dispersion, featuring a double-minimum structure, and the two-body
interaction, these systems possess a complex phase diagram. Three different quantum phases can be observed, i.e., a
stripe, a plane-wave, and a single-minimum phase, each characterized by different broken symmetries. The condensate
dynamics is also significantly affected by the spin-orbit coupling, as revealed by the behavior of the Bogoliubov
spectrum in infinite systems and the behavior of the discretized collective mode frequencies in trapped configurations,
especially close to the phase transitions. In turn, the superfluid and rotational properties are also deeply modified by
the coupling between the motional and spin degree of freedom. Finally, a special attention is devoted to the stripe phase
and its supersolid character, which can be clearly revealed by the study of its dynamic features.
\end{abstract}

\section{Introduction}
\label{sec:intro}
The possibility to realize artificial gauge fields has opened novel exciting perspectives in the field of ultracold atomic
gases~\cite{Dalibard2011review,Goldman2014review}. Among the various implementations, the use of Raman coupling has proven
very successful in generating spin-orbit coupling (SOC) in Bose-Einstein condensates (BECs). After the seminal papers by
Spielman’s team~\cite{Lin2009,Lin2011}, the experimental and theoretical research activity in this field has significantly
grown in the last decade, providing a better understanding of the novel phase diagram characterizing these systems and of
their equilibrium, dynamic and superfluid properties (see the reviews~\cite{Galitski2013review,Zhou2013review,Zhai2015review,
Li2015review,Zhang2016review,Recati2022review,Martone2023review} and references therein). Bose-Einstein condensates with
Raman-induced SOC exhibit a variety of quantum phases, including the so-called single-minimum, plane-wave, and stripe phases.
These can be obtained by properly tuning the Raman coupling, which is fixed by the intensity of the laser beams providing
the relevant transitions between the hyperfine states of the employed atomic species, and the momentum transferred to the
system. While the uniform single-minimum and plane-wave phases have been already the object of rather systematic experimental
work, experiments on the intriguing stripe phase have only recently become available~\cite{Li2017}, opening new perspectives
in the study of supersolidity. This is a non-intuitive phenomenon characterized by the spontaneous and simultaneous breaking
of two continuous symmetries: the breaking of phase symmetry, yielding the effects of superfluidity, and the breaking of
translation invariance, responsible for the crystal behavior of the system.

In the present lecture notes we will summarize some of the most relevant theoretical and experimental features characterizing
BECs with one-dimensional SOC involving two hyperfine states, giving rise to effective spin-$1/2$ configurations. Special
emphasis will be given to illustrate the nature of the quantum phases exhibited by these systems and their dynamic and
superfluid behavior.

\section{Single-particle Hamiltonian}
\label{sec:sp}
The pioneering experiment by Lin \textit{et al.} reported in Ref.~\cite{Lin2011} was performed on an ultracold gas of $^{87}$Rb
atoms in the $F = 1$ hyperfine manifold. The degeneracy of the levels within the manifold was lifted by the presence of a bias
magnetic field. In order to achieve SOC, the authors employed a pair of Raman lasers, which induced transitions between the three
levels by transferring momentum $2 \hbar \vec{k}_R$ and energy $\hbar \Delta \omega_L$ to the gas. $\Delta \omega_L$ corresponds
to the frequency difference between the two lasers, which in Ref.~\cite{Lin2011} is tuned close to resonance with the transition
between the $|F=1,m_F=-1\rangle \equiv |\downarrow\rangle$ and $|F=1,m_F=0\rangle \equiv |\uparrow\rangle$ levels, characterized
by a frequency $\omega_Z$ fixed by the external magnetic field. This enables one to describe the system using an effective
spin-$1/2$ model: indeed, the laser field is far detuned from the transition between $|F=1,m_F=0\rangle$ and $|F=1,m_F=+1\rangle$,
whose frequency is significantly larger than $\omega_Z$ owing to the quadratic Zeeman effect, hence, the latter level can be
adiabatically eliminated. The resulting Hamiltonian, which depends on the phase of the Raman laser field, can be made space- and
time-independent by performing the spin rotation $\mathcal{U} = \exp \left[ \mi (2 k_R x - \Delta\omega_L t) \sigma_z / 2
\right]$~\cite{Lin2011,Martone2012}. This yields the final expression
\begin{equation}
h_{\mathrm{SO}} = \frac{\left( \vec{p} - \hbar \vec{k}_R \sigma_z \right)^2}{2m} + \frac{\hbar\Omega_R}{2} \, \sigma_x
+ \frac{\hbar \delta_R}{2} \, \sigma_z + V_{\mathrm{ext}}(\vec{r}) \, ,
\label{eq:h_SO}
\end{equation}
where $m$ denotes the atom mass, $\sigma_{x,y,z}$ the $2 \times 2$ Pauli matrices, and $V_{\mathrm{ext}}(\vec{r})$ the external
trapping potential. We take $\vec{k}_R = k_R \hat{\vec{e}}_x$ along the positive $x$ direction, $\hat{\vec{e}}_x$ being the
corresponding unit vector, and introduce the energy scale $E_R = (\hbar k_R)^2 / 2m$. The Raman coupling $\Omega_R$ can be tuned
by changing the intensity of the laser beams, whose detuning from Raman resonance is quantified by $\delta_R \equiv \Delta \omega_L
- \omega_Z$. The Hamiltonian~\eqref{eq:h_SO} features a one-dimensional SOC, given by a superposition of the Rashba~\cite{Bychkov1984}
and Dresselhaus~\cite{Dresselhaus1955} terms with equal weights.

\begin{figure}
\centering
\includegraphics{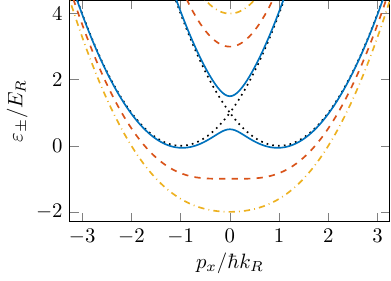}
\caption{Single-particle energy dispersion law~\eqref{eq:sp_en} as a function of canonical momentum. Both the lower and upper
branches are plotted, taking zero $\delta_R$ and $\hbar\Omega_R / E_R = 1.0$ (blue solid line), $4.0$ (red dashed line), and
$6.0$ (yellow dash-dotted line). The bare dispersion at vanishing Raman coupling $\Omega_R$ is also shown (black dotted line).
Reprinted from~\cite{Martone2023review} under the terms of the Creative Commons Attribution 4.0 International License (see
\texttt{http://creativecommons.org/licenses/by/4.0/}).}
\label{fig:sp_spectrum}
\end{figure}

If $V_{\mathrm{ext}}(\vec{r}) = 0$, the Hamiltonian~\eqref{eq:h_SO} commutes with the canonical momentum operator $\vec{p} =
- \mi \hbar \nabla$ (see also Sec.~\ref{sec:symm} below). By diagonalizing this Hamiltonian in momentum space one gets the energy
dispersion law
\begin{equation}
\varepsilon_\pm(\vec{p})
= \frac{\vec{p}^2}{2m} + E_R \pm \frac{\hbar}{2} \sqrt{\left( \frac{2 k_R p_x}{m} - \delta_R \right)^2 + \Omega_R^2} \, .
\label{eq:sp_en}
\end{equation}
It is comprised of two branches, a lower and an upper one, separated by a gap equal, for $\delta_R = 0$, to $\hbar\Omega_R$.
These branches are shown in Fig.~\ref{fig:sp_spectrum} for several values of the Raman coupling $\Omega_R$ and at zero detuning
$\delta_R$. Notice that at small $\Omega_R$, the lower branch possesses two degenerate global minima located at $\vec{p} =
\pm \hbar k_1^{(0)} \hat{\vec{e}}_x$, where
\begin{equation}
k_1^{(0)} = k_R \sqrt{1 - \left(\frac{\hbar \Omega_R}{4 E_R}\right)^2} \, .
\label{eq:sp_k1}
\end{equation}
This double-minimum structure exists as long as $\hbar\Omega_R < 4 E_R$, while for larger Raman coupling the two minima coalesce
into a single one at $\vec{p} = 0$. In the following, we will highlight some implications of this peculiar single-particle dispersion
on the behavior of the gas in the presence of interparticle interactions.

\section{Broken symmetries of the spin-orbit Hamiltonian}
\label{sec:symm}
The spin-orbit Hamiltonian~\eqref{eq:h_SO} breaks important symmetries, giving rise to novel many-body features which will be discussed
in the following sections and that it is worth anticipating here. 

\paragraph{Breaking of Galilean invariance}
In the absence of the external potential $V_{\mathrm{ext}}$, the Hamiltonian is translation invariant, $[h_{\mathrm{SO}},\vec{p}] = 0$
(although translational symmetry can be spontaneously broken giving rise to a supersolid configuration, see Sec.~\ref{sec:stripe}).
However, due to SOC, the Hamiltonian breaks Galilean invariance. This can be understood as follows. The spin-orbit Hamiltonian in a frame
moving along $x$ with velocity $v$, obtained by applying the unitary Galilean transformation $\exp(\mi m v x / \hbar)$ to the original
Hamiltonian~\eqref{eq:h_SO}, takes the form 
\begin{equation}
(h_{\mathrm{SO}})'= h_{\mathrm{SO}} + \frac{m v^2}{2} - m v P_x \, .
\label{eq:h_SO_boost}
\end{equation}
This expression does not commute with $h_{\mathrm{SO}}$ owing to the presence of the spin term in the $x$ component of the physical
momentum operator $\vec{P} = \vec{p} - \hbar \vec{k}_R \sigma_z$ (that fixes the velocity of particles), which yields the commutation
relation 
\begin{equation}
[h_{\mathrm{SO}},P_x]= \mi \hbar^2 k_R \Omega_R \sigma_y \, .
\label{eq:h_SO_comm}
\end{equation}
The breaking of Galilean invariance is responsible for important consequences concerning the Landau criterion for
superfluidity~\cite{Zhu2012,Zheng2013} and the stability of supercurrents~\cite{Ozawa2013}. As explicitly discussed in
Sec.~\ref{sec:sup_rot}, the lack of Galilean invariance dramatically also affects the value of the superfluid density,
even if the system is translationally invariant and the density is uniform.

\paragraph{Parity and Time reversal symmetries}
The spin-orbit Hamiltonian~\eqref{eq:h_SO} additionally breaks both parity ($\mathcal{P}$) and time reversal ($\mathcal{T}$) symmetries.
However, if $\delta_R = 0$, it is invariant under the two $\mathbb{Z}_2$ symmetries represented by the operators $\sigma_x \mathcal{P}$
and $\sigma_z \mathcal{T}$. The former (latter) performs a rotation by $\pi$ along the $x$ ($z$) spin axis followed by a parity (time
reversal) operation. Invariance under these two discrete symmetries is retained also in the presence of two-body contact interaction, 
provided that the two intraspecies scattering lengths are equal (see next section); notice that if this condition is not fulfilled,
or if $\delta_R \neq 0$, the many-body Hamiltonian is not invariant under $\sigma_x \mathcal{P}$ and $\sigma_z \mathcal{T}$ separately,
but it is invariant under their product. In addition, these two symmetries are spontaneously broken in the plane-wave phase (see next
section), resulting in the breaking of the usual identity $\omega(-\vec{q}) = \omega(\vec{q})$ characterizing the excitation spectrum,
as discussed in Sec.~\ref{sec:hd}.

\section{Interacting ground state}
\label{sec:gs}
Let us now assume that the $N$ constituents of our spin-orbit-coupled BEC, enclosed in a volume $V$, interact through a two-body
contact potential. Within the Gross-Pitaevskii (GP) mean-field theory, this many-body system can be described by a two-component
spinor wave function $\Psi$. We take $\Psi$ normalized to the total number of particles, i.e., $\int_V \dif\vec{r} \, \Psi^\dagger
\Psi = N$. The energy of the system as a functional of $\Psi$ reads
\begin{equation}
E[\Psi] = \int_V \dif\vec{r} \left( \Psi^\dagger h_{\mathrm{SO}} \Psi + \frac{g_{dd}}{2} n^2
+ \frac{g_{ss}}{2} s_z^2 + g_{ds} n s_z \right) \, .
\label{eq:gp_en}
\end{equation}
Here, $n = \Psi^\dagger \Psi$ is the total density of the condensate, with average value $\bar{n} = N/V$, and $s_z =
\Psi^\dagger \sigma_z \Psi$ is the spin density. The three quantities $g_{dd} = (g_{\uparrow\uparrow} + g_{\downarrow\downarrow}
+ 2 g_{\uparrow\downarrow}) / 4$, $g_{ss} = (g_{\uparrow\uparrow} + g_{\downarrow\downarrow} - 2 g_{\uparrow\downarrow}) / 4$,
and $g_{ds} = (g_{\uparrow\uparrow} - g_{\downarrow\downarrow}) / 4$ represent the density-density, spin-spin, and density-spin
coupling constants, respectively. They are linear combinations of the couplings $g_{\sigma\sigma'} = 4 \pi \hbar^2
a_{\sigma\sigma'} / m$ ($\sigma,\sigma' = \uparrow,\downarrow$) in the various spin channels, with $a_{\sigma\sigma'}$ the
corresponding scattering lengths.

The wave function $\Psi_0$ of the condensate at equilibrium is found from the stationarity condition $\delta E' = 0$ of the
grand-canonical energy $E'[\Psi_0] = E[\Psi_0] - \mu \int_V \dif\vec{r} \, \Psi_0^\dagger \Psi_0$, $\mu$ being the chemical
potential. This yields the time-independent Gross-Pitaevskii equation~\cite{Pitaevskii_Stringari_book}
\begin{equation}
[ h_\mathrm{SO} + g_{dd} (\Psi_0^\dagger \Psi_0)
+ g_{ss} (\Psi_0^\dagger \sigma_z \Psi_0) \sigma_z ] \Psi_0 = \mu \Psi_0 \, .
\label{eq:ti_gp_eq}
\end{equation}
Here and henceforth, we take zero Raman detuning $\delta_R$ and equal intraspecies couplings ($g_{ds} = 0$), unless otherwise
specified. The ground state of the BEC corresponds to the solution of Eq.~\eqref{eq:ti_gp_eq} having the lowest value of the
energy~\eqref{eq:gp_en}. In the case of an infinite system, i.e., $V_{\mathrm{ext}} = 0$, the following Ansatz has been
employed~\cite{Ho2011,Li2012a} (see also~\cite{Wang2010,Wu2011} for similar proposals in BECs with Rashba SOC):
\begin{equation}
\Psi_0(\vec{r}) = \sqrt{\bar{n}} \left[ C_+
\begin{pmatrix}
\cos\vartheta \\
- \sin\vartheta
\end{pmatrix}
\me^{\mi k_1 x}
+ C_-
\begin{pmatrix}
- \sin\vartheta \\
\cos\vartheta
\end{pmatrix}
\me^{- \mi k_1 x}
\right] .
\label{eq:ansatz}
\end{equation}
This expression represents a superposition of two counterpropagating plane waves, with complex weights $C_\pm$ obeying the
normalization condition $|C_+|^2 + |C_-|^2 = 1$. The two plane waves are multiplied by a real spinor, whose spin polarization,
quantified by $0 \leq \vartheta \leq \pi/4$, is related to the wave vector by the constraint $k_1 = k_R \cos 2\vartheta$
imposed by energy minimization, as explained below.

In a noninteracting setup, where $g_{dd} = g_{ss} = 0$, one can easily check that the Ansatz~\eqref{eq:ansatz} is an exact
solution of the GP equation~\eqref{eq:ti_gp_eq} and minimizes the energy~\eqref{eq:gp_en} if $k_1$ is taken equal to the
single-particle value~\eqref{eq:sp_k1}; on the other hand, the energy is independent of the value of $C_\pm$, which remains
arbitrary. One can expect that the mean-field ground state of a dilute and weakly interacting spin-orbit-coupled BEC should
still be of the kind~\eqref{eq:ansatz}, although the values of $C_\pm$, $\vartheta$, and $k_1$ will now depend on the
interaction. It turns out that the wave function~\eqref{eq:ansatz} is not a solution of the GP equation~\eqref{eq:ti_gp_eq}
in the presence of interaction, unless either $C_+$ or $C_-$ is zero. However, this Ansatz is able to capture the most
relevant features of the BEC ground state and represents a fairly accurate approximation to the exact wave function if one
chooses the parameters so as the total energy~\eqref{eq:gp_en} is minimized. This variational procedure yields the phase
diagram of Fig.~\ref{fig:phase_diag}, which features three quantum phases~\cite{Ho2011,Li2012a} with well-defined occupations
$|C_\pm|^2$ of the two plane-wave components of Eq.~\eqref{eq:ansatz}, in stark contrast with the ideal gas case. Taking
$g_{ss} > 0$ and a low average density $\bar{n}$, the phases that one observes at increasing Raman coupling $\Omega_R$ are
the following:
\begin{figure}
\centering
\includegraphics{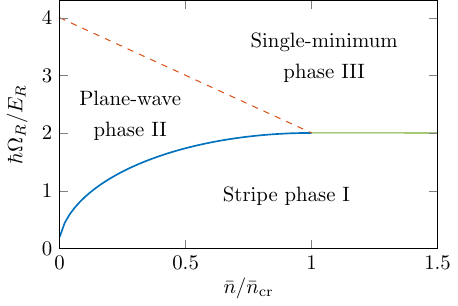}
\caption{Raman coupling vs. average density phase diagram of a BEC with SOC. The solid blue and green lines indicate the
first-order transition from the stripe to the PW and SM phase, respectively. The dashed red line shows the second-order
PW--to--SM phase transition. Adapted from~\cite{Martone2012}.}
\label{fig:phase_diag}
\end{figure}
\begin{enumerate}
\renewcommand{\labelenumi}{\Roman{enumi}.}
\item At small $\Omega_R$, the system is in the stripe phase, also called the phase-mixed configuration because the two momentum
components of the wave function~\eqref{eq:ansatz} have equal occupations $|C_+|^2 = |C_-|^2 = 1 / 2$ and coexist in the whole
space. This phase has zero spin polarization $\langle \sigma_z \rangle = \int_V \dif\vec{r} \, s_z(\vec{r})$. The condensate
density is characterized by periodic fringes of wavelength $\pi / k_1$, where $k_1 = k_R \sqrt{1 - [\hbar \Omega_R / (4 E_R
+ g_{dd} \bar{n})]^2}$. These fringes result from the interference of the two plane-wave components of Eq.~\eqref{eq:ansatz}
and are revealing of a spontaneous breaking mechanism of translation symmetry, which connects the stripe phase to the phenomenon
of supersolidity, see Sec.~\ref{sec:stripe}. It is worth mentioning that the full condensate wave function in the stripe phase
includes higher harmonic components oscillating at wave vector $\pm 3 k_1, \pm 5 k_1, \ldots$~\cite{Li2013,Martone2021}, which
arise because of the nonlinear terms of the Gross-Pitaevskii equation~\eqref{eq:ti_gp_eq} and are neglected in the variational
Ansatz~\eqref{eq:ansatz}. An alternative approach for studying this phase, yielding analytic and quantitatively more accurate
results than Eq.~\eqref{eq:ansatz}, relies on a perturbation analysis of the wave function and the various observables at low
Raman coupling~\cite{Martone2021};
\item By increasing $\Omega_R$, one can induce a first-order phase transition to the so-called plane-wave (PW) phase. The
critical Raman coupling characterizing this transition is density-dependent, and for $\bar{n} \to 0$ takes the simple
expression~\cite{Ho2011,Li2012a}
\begin{equation}
\hbar\Omega_{\mathrm{cr1}} = 4 E_R \sqrt{\frac{2 g_{ss}}{g_{dd} + 2 g_{ss}}} \, .
\label{eq:omega_cr1}
\end{equation}
In the PW phase, energy minimization imposes that either $C_-$ or $C_+$ vanishes, meaning that only one of the two momentum
states entering Eq.~\eqref{eq:ansatz} is occupied. This configuration spontaneously breaks the $\sigma_x \mathcal{P}$ and
$\sigma_z \mathcal{T}$ discrete symmetries introduced in Sec.~\ref{sec:symm} and is thus twofold degenerate. The two physically
distinct ground states associated with the PW phase are characterized by opposite values of the momentum, $\vec{p} = \pm \hbar
k_1 \hat{\vec{e}}_x$, and of the spin polarization, $\langle \sigma_z \rangle = \pm N k_1 / k_R$, where $k_1 = k_R \sqrt{1 -
[\hbar \Omega_R / (4 E_R - 2 g_{ss} \bar{n})]^2}$. Unlike the stripe phase, the density in the two PW states is uniform;
\item By further increasing $\Omega_R$, one observes a second-order phase transition at the critical value
\begin{equation}
\hbar\Omega_{\mathrm{cr2}} = 4 E_R - 2 g_{ss} \bar{n} \, .
\label{eq:omega_cr2}
\end{equation}
Above this threshold, the momentum and the spin polarization vanish, and the system enters another uniform configuration, the
single-minimum (SM) phase. Experimentally, the PW--to--SM transition has been detected from measurements of the momentum
distribution~\cite{Lin2011} and of the spin polarization~\cite{Hamner2014} as functions of $\Omega_R$.
\end{enumerate}

The above scenario is valid for densities smaller than $\bar{n}_{\mathrm{cr}} = E_R g_{dd} / [g_{ss} (g_{dd} + g_{ss})]$, while above
this critical value, the PW phase is no longer available, and one observes a direct first-order transition from the stripe to the SM
phase. As shown in Fig.~\ref{fig:phase_diag}, the two regimes are separated by a tricritical point, where all the three phases can
coexist. Notice that the studies~\cite{Ho2011,Li2012a} rely on the mean-field approximation, whose validity in dilute three-dimensional
BECs with one-dimensional SOC is confirmed by the smallness of the quantum depletion~\cite{Zheng2013,Chen2018}. Interestingly,
subsequent Monte Carlo analyses have shown that many of the above features persist even at stronger interactions~\cite{Sanchez2020},
and that interatomic correlations actually tend to stabilize even more the stripe phase, resulting in a significant reduction of the
value of $\bar{n}_{\mathrm{cr}}$. On the other hand, first theoretical~\cite{Yu2014} and experimental~\cite{Ji2014} works at finite
temperature have shown that thermal effects seem to favor the PW over the stripe phase, especially reducing the critical Raman
coupling separating the two configurations with respect to the zero-temperature value~\eqref{eq:omega_cr1}. We finally mention that
the phase diagram for $g_{ss} < 0$ still features the PW and SM phases as well as the second-order transition at the critical Raman
coupling~\eqref{eq:omega_cr2}, while the stripe phase is absent, since it is not energetically favorable in these conditions.

A key quantity for characterizing the static and dynamic properties of spin-orbit-coupled BECs is the magnetic susceptibility. It can
be computed by adding a weak external perturbation $- h \sigma_z$ to the single-particle Hamiltonian~\eqref{eq:h_SO}, and by evaluating
the subsequent response $\chi_M = \lim_{h \to 0} \dif \langle \sigma_z \rangle / \dif h$ of the spin polarization. The magnetic
susceptibility in the PW and SM phase is given, respectively, by~\cite{Li2012b}
\begin{align}
\chi_M^{(\mathrm{II})} &{} =
\frac{2 \Omega_R^2}{\hbar\Omega_{\mathrm{cr2}} \left(\Omega_{\mathrm{cr2}}^2 - \Omega_R^2\right)} \, ,
\label{eq:chi_pw} \\
\chi_M^{(\mathrm{III})} &{} = \frac{2}{\hbar\left(\Omega_R - \Omega_{\mathrm{cr2}}\right)} \, .
\label{eq:chi_sm}
\end{align}
Notice the divergence of $\chi_M$ at the PW--to--SM phase transition, confirming once more its second-order nature. The magnetic
susceptibility was experimentally determined from the analysis of the center-of-mass oscillation in~\cite{Zhang2012} (see also
Sec.~\ref{sec:hd} below) and found in good agreement with the prediction \eqref{eq:chi_pw}--\eqref{eq:chi_sm}. The role of the
magnetic susceptibility is better understood in the $g_{ss} = 0$ case, where the condensate wave function remains the same as in
the ideal gas model (see Sec.~\ref{sec:sp}). The curvature of the lower branch of the single-particle spectrum~\eqref{eq:sp_en}
close to the minima defines an effective mass $1/m^* = \partial^2 \varepsilon_- / \partial p_x^2|_{\vec{p} = \pm \hbar k_1
\hat{\vec{e}}_x}$, which is related to $\chi_M$ as
\begin{equation}
\frac{m^*}{m} = 1 + 2 E_R \chi_M \, .
\label{eq:eff_mass}
\end{equation}
Notice the divergence of $m^*$ at the critical point $\Omega_{\mathrm{cr2}}$ (equal to $4 E_R$ for $g_{ss} = 0$), where the
two degenerate minima of the single-particle spectrum~\eqref{eq:sp_en} merge into a single one, signaling the peculiar $p_x^4$
behavior of the dispersion at small momenta.

The transition between the PW and the SM phase discussed above, and in particular the divergent behavior of the magnetic
susceptibility (see Eqs.~\eqref{eq:chi_pw} and~\eqref{eq:chi_sm}), shares a close analogy with the ferromagnetic-to-paramagnetic
transition exhibited by Rabi-coupled Bose-Einstein condensed mixtures (see, for example, \cite{Recati2022review}). Notice, however,
that in the case of Rabi coupling, where the momentum transfer $2 \hbar \vec{k}_R$ can be set equal to zero, the phase transition
exists only for negative values of the spin-spin coupling constant $g_{ss}$, and the value of the coupling at the transition is
uniquely fixed by interaction effects ($\hbar\Omega_{\mathrm{cr}} = - 2 g_{ss} \bar{n}$). Similar physics can also be observed in
cold atomic gases in shaken optical lattices~\cite{Parker2013,Ha2015}, where by varying the amplitude of the lattice shaking, the
energy dispersion in momentum space smoothly changes from single-minimum to double-minimum. Thinking of these two minima as two
pseudospin states, the ferromagnetic interaction present in these systems then favors the occurrence of a magnetic-like
configuration, with the possible formation of multiple ferromagnetic domains~\cite{Parker2013}. We finally mention that, as shown in
Ref.~\cite{Hamner2014}, an interacting spin-orbit-coupled BEC in a uniform phase can be mapped to the Dicke model of quantum optics,
describing the interaction of a set of two-level atoms with a single-mode optical cavity. In this analogy, the PW--to--SM transition
of a BEC with SOC plays the role of the transition between the superradiant and the normal phase occurring in the Dicke model.

\section{Collective modes and hydrodynamic theory}
\label{sec:hd}
Bose-Einstein condensates with SOC possess interesting dynamic properties, which are a direct consequence of their peculiar phase
diagram. In this section, we focus the attention on the PW and SM phases in both infinite and trapped configurations. The nature of
the collective oscillations in the stripe phase will be discussed in Sec.~\ref{sec:stripe}.

In order to study how the condensate wave function $\Psi$ evolves in time, one has to solve the time-dependent Gross-Pitaevskii
equation~\cite{Pitaevskii_Stringari_book}
\begin{equation}
\mi \hbar \partial_t \Psi =
\left[ h_\mathrm{SO} + g_{dd} \left(\Psi^\dagger \Psi\right)
+ g_{ss} \left(\Psi^\dagger \sigma_z \Psi\right) \sigma_z \right] \Psi \, .
\label{eq:td_gp_eq}
\end{equation}
This equation stems from the stationarity condition $\delta A_{\mathrm{GP}} = 0$ of the Gross-Pitaevskii action $A_{\mathrm{GP}}
= \int \dif t \int \dif\vec{r} \left( \mi \hbar \Psi^\dagger \partial_t \Psi - E[\Psi] \right)$. Notice that for stationary
configurations, whose wave function evolves in time according to $\Psi(\vec{r},t) = \me^{- \mi \mu t / \hbar} \Psi_0(\vec{r})$,
it reduces to the time-independent equation~\eqref{eq:ti_gp_eq}. In the case of small deviations from a given equilibrium state,
one can write $\Psi(\vec{r},t) = \me^{- \mi \mu t / \hbar} \left[ \Psi_0(\vec{r}) + \delta \Psi(\vec{r},t) \right]$ and use this
Ansatz to linearize Eq.~\eqref{eq:td_gp_eq} and compute the fluctuation term $\delta \Psi$. We recall that the linearized
Gross-Pitaevskii equation has solutions of the kind $\delta \Psi(\vec{r},t) = U(\vec{r}) \me^{- \mi \omega t} + V^*(\vec{r})
\me^{\mi \omega t}$, characterized by an oscillation frequency $\omega$ and two complex spinor amplitudes $U(\vec{r})$ and
$V(\vec{r})$, which correspond to the Bogoliubov modes of the BEC~\cite{Pitaevskii_Stringari_book}; in general, $\delta \Psi$
can be expressed as a linear superposition of these modes.

\begin{figure}
\centering
\includegraphics[scale=0.7]{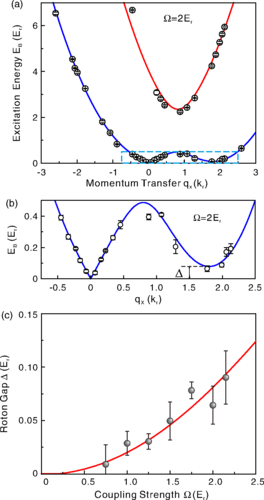}
\caption{(a) Theoretical prediction for the lower (blue solid line) and upper (red solid line) branch of the Bogoliubov spectrum
of a spin-orbit-coupled BEC in the PW phase with negative momentum. The circles represent the measured values. (b) Zoom on the
bottom part of (a), where the roton-maxon structure in the lower branch is visible. (c) Predicted (solid line) and measured
(circles) value of the roton gap as a function of the Raman coupling, showing its softening as $\Omega_R$ approaches the critical
value~\eqref{eq:omega_cr1}. From~\cite{Ji2015}. Reprinted figure with permission from Si-Cong Ji \textit{et al.}, \textit{Physical
Review Letters} \textbf{114}, 105301 (2015). Copyright 2015 by the American Physical Society.}
\label{fig:pw_spectrum}
\end{figure}

For an infinite condensate in the PW and SM phases, the Bogoliubov amplitudes $U(\vec{r})$ and $V(\vec{r})$ are plane waves.
Hence, one can label each excitation by its momentum $\hbar\vec{q}$ with respect to the condensate and compute the corresponding
frequency~\cite{Zheng2012,Martone2012,Zheng2013}. As shown in Fig.~\ref{fig:pw_spectrum}, the dispersion of elementary excitations
is made of two branches, which are reminiscent of those of the single-particle spectrum~\eqref{eq:sp_en}. The lower branch is
gapless and displays a phonon-like behavior at long wavelengths, $\omega_-(\vec{q}) \simeq c(\theta_{\vec{q}}) q$. Different
from standard BECs, the sound velocity is anisotropic as it depends on the angle $\theta_{\vec{q}}$ between the momentum $\vec{q}$
and the positive $x$ direction. In addition, in the PW phase, the velocities of sound waves propagating along opposite directions
in space are different (unless $g_{ss} = 0$) because of the spontaneous breaking of the two $\mathbb{Z}_2$ symmetries introduced
in Sec.~\ref{sec:symm}. In both the PW and SM phases, the sound velocities $c_{x,+}$ and $c_{x,-}$, relative to the propagation of
sound parallel and anti-parallel to the $x$ axis, obey the relevant relation~\cite{Martone2012}
\begin{equation}
m c_{x,+} c_{x,-} = \frac{\kappa^{-1}}{1 + 2 E_R \chi_M} \, ,
\label{eq:sound_chi}
\end{equation}
where $\kappa^{-1} = \bar{n} (\partial \mu / \partial \bar{n})$ is the inverse compressibility of the gas and $c_{x,+} = c_{x,-}$
in the SM phase. The sound velocities $c_{x,+}$ and $c_{x,-}$ were actually measured in both the PW and SM phases~\cite{Ji2015},
confirming their strong suppression at the critical point~\eqref{eq:omega_cr2} separating the two phases as a consequence of the
divergence of the magnetic susceptibility~\eqref{eq:chi_pw}--\eqref{eq:chi_sm}. Conversely, along the transverse directions, the
sound velocity is given by the standard relation $m c_\perp^2 = \kappa^{-1}$.

As visible in Fig.~\ref{fig:pw_spectrum}, at larger wave vector size, the Bogoliubov spectrum in the PW phase exhibits a peculiar
roton-maxon structure. Physically, the presence of a roton minimum at $\vec{q} \sim 2 k_1 \hat{\vec{e}}_x$ in the spectrum of a
plane-wave condensate with negative momentum (or at $\vec{q} \sim - 2 k_1 \hat{\vec{e}}_x$ for a positive-momentum plane-wave BEC)
means that atoms can be transferred to the degenerate empty ground state at small energetic cost, corresponding to the roton gap.
This gap is finite only in the presence of interaction, and it reduces as one gets closer to the transition to the stripe phase,
thus providing the onset for the occurrence of a crystalline structure. The roton-maxon structure was detected experimentally in
Refs.~\cite{Khamehchi2014,Ji2015}, in excellent agreement with previous theoretical predictions~\cite{Zheng2012,Martone2012,
Zheng2013}, see Fig.~\ref{fig:pw_spectrum}; a similar behavior was also found in BECs in shaken optical lattices~\cite{Ha2015}.
The presence of this structure on only one side of the Bogoliubov spectrum is another consequence of the violation of the relation
$\omega(-\vec{q}) = \omega(\vec{q})$, resulting from the spontaneous violation of the $\mathbb{Z}_2$ symmetries discussed in
Sec.~\ref{sec:symm}. We finally mention that the spectrum also includes an upper gapped branch, whose frequency separation from the
lower one is of the order of the Raman coupling $\Omega_R$, as in the noninteracting model of Sec.~\ref{sec:sp}.

A relation similar to Eq.~\eqref{eq:sound_chi} holds for the frequencies of the collective modes in finite systems confined by a
harmonic potential $V_{\mathrm{ext}}(\vec{r}) = m(\omega_x^2 x^2 + \omega_y^2 y^2 + \omega_z^2 z^2) / 2$, $\omega_{x,y,z}$ being
the trap frequencies. In the case of the dipole mode, this stems from the commutator $[h_{\mathrm{SO}},x] = - \mi \hbar P_x / m$
of the Hamiltonian~\eqref{eq:h_SO} with the dipole operator, featuring the physical momentum $P_x = p_x - \hbar k_R \sigma_z$ along
$x$, which is modified by SOC. By means of sum-rule techniques, one finds the following upper bound for the dipole oscillation
frequency~\cite{Li2012b}:
\begin{equation}
\omega_{\mathrm{D}} = \frac{\omega_x}{1 + 2 E_R \chi_M} \, .
\label{eq:dipole_chi}
\end{equation}
The validity of this estimate has been confirmed in the experiment of Ref.~\cite{Zhang2012}. Later works have predicted a similar
behavior for the frequency of the breathing mode~\cite{Chen2017,Geier2021} (see also Fig.~\ref{fig:trap_modes} below). 

The macroscopic excitations of the system, corresponding to small wave vectors $\vec{q}$ and frequencies $\omega$ can be conveniently
described by employing the hydrodynamic formalism where one ignores quantum pressure effects. This formalism was first applied to the case
of spin-orbit-coupled BECs in~\cite{Martone2012} and is characterized by an important simplification of the behavior of the relative phase
between the two spin components, which is locked if the frequency of the elementary excitations satisfies the condition $\omega(\vec{q})
\ll \Omega_R$.   

In the hydrodynamic regime, the energy associated with the Raman-induced spin-orbit Hamiltonian then depends on three variables: the total
density $n = n_\uparrow + n_\downarrow$, the spin density $s_z = n_\uparrow - n_\downarrow$, and the phase $\phi$ of the order parameter.
One finds $E = \int \dif\vec{r} \, \epsilon(\vec{r})$, with
\begin{equation}
\epsilon(\vec{r}) = \frac{\hbar^2}{2m} \left(\nabla\phi\right)^2 n - \frac{\hbar k_R}{m} \left(\nabla_x\phi\right) s_z + \frac{g n^2}{2}
+ V_{\mathrm{ext}} n - \frac{\hbar \Omega_R}{2} n + \frac{\hbar\Omega_R}{4} \frac{s_z^2}{n} \, .
\label{eq:hd_en} 
\end{equation}
For simplicity, here we have assumed $g_{ss} = 0$ and considered the SM phase where, at equilibrium, $\nabla \phi = 0$ and $s_z = 0$.
The extension of the formalism to the PW phase is straightforward~\cite{Martone2012}.

The time-dependent hydrodynamic equations of motion are obtained by applying the variational procedure $\delta A = 0$ to the action $A =
\int \dif t \int \dif\vec{r} \left[ \epsilon(\vec{r}) + \hbar n \partial_t \phi \right]$. Variation of the action with respect to
the phase $\phi$ yields the equation of continuity
\begin{equation}
\partial_t n + \frac{\hbar}{m} \nabla \cdot (n \nabla\phi) - \frac{\hbar k_R}{m} \nabla_x s_z = 0 \, ,
\label{eq:hd_cont}
\end{equation}
where one recognizes the spin contribution that modifies the definition of the current density along the $x$ direction according to
$j_x = (\hbar/m) n \nabla_x \phi - (\hbar k_R/m) s_z$, which is no longer uniquely fixed by the gradient of the phase of the order parameter,
as happens in usual BECs. The spin contribution to the current actually can cause the violation of the irrotational
nature of the velocity field $\vec{v} = \vec{j} / n$, as we will discuss in Sec.~\ref{sec:sup_rot}. 

Variation of $A$ with respect to the total density yields the equation for the phase 
\begin{equation}
\hbar \partial_t \phi = - \frac{\hbar\Omega_R}{2} + g n + V_{\mathrm{ext}} \, ,
\label{eq:hd_phase}
\end{equation}
where we have ignored terms quadratic in the gradient of the phase and in $s_z$. Finally, variation with respect to the spin density yields
the novel hydrodynamic equation
\begin{equation}
- \frac{\hbar k_R}{m} \nabla_x \phi + \frac{\Omega_R}{2} \frac{s_z}{n} = 0 \, ,
\label{eq:hd_novel}
\end{equation}
which fixes a nontrivial relationship between the gradient of the phase and the spin density. By releasing the condition
$\phi_{\mathrm{rel}} \equiv \phi_\uparrow - \phi_\downarrow = 0$, this latter equation would contain the time derivative of the relative
phase $\phi_{\mathrm{rel}}$.

The above relations form a self-consistent set of equations accounting for the dynamics of a spin-orbit-coupled Bose gas. They can be also used,
with the proper inclusion of an external perturbation, to study the dynamic response to slowly varying macroscopic perturbations of density or
spin nature. In the limit of small-amplitude oscillations, they reduce to a single equation for the density fluctuations,
\begin{equation}
m \partial_t^2 \delta n - \nabla_\perp \cdot \left(g n \nabla_\perp \delta n\right)
- \frac{m}{m^*} \nabla_x \cdot \left(g n \nabla_x \delta n\right) = 0 \, ,
\label{eq:hd_small}
\end{equation}
and in this form, the equation holds in both the SM and the PW phases, with the proper inclusion of the effective mass~\eqref{eq:eff_mass}
and the magnetic susceptibility given, respectively, by Eq.~\eqref{eq:chi_sm} and~\eqref{eq:chi_pw}. Equation~\eqref{eq:hd_small} provides the
simple expression $c = \sqrt{g n / m^*}$ for the sound velocity in uniform matter along the $x$ axis and reproduces the result~\eqref{eq:dipole_chi}
for the frequency of the dipole oscillation in the presence of harmonic trapping.

\section{Superfluidity and rotational properties}
\label{sec:sup_rot}
As discussed in Sec.~\ref{sec:symm}, the spin-orbit Hamiltonian~\eqref{eq:h_SO} is invariant under translations but not under Galilean
transformations, as its commutator~\eqref{eq:h_SO_comm} with the $x$ component of the physical momentum is nonvanishing. This deeply affects
the superfluid and rotational properties of the condensate.

We will follow Baym's approach to superfluidity, which is based on the definition
\begin{equation}
\frac{\rho_n}{\rho} = \frac{1}{N m} \lim_{q_y \to 0}
\left[ \sum_{n \neq 0} \frac{\left|\langle 0 | J^T_x(q_y) | n \rangle\right|^2}{E_n - E_0} + \left( q_y \to - q_y \right) \right]
\label{eq:norm_dens}
\end{equation}
for the normal density evaluated along the $x$ direction, with $\rho = m \bar{n}$ the total mass density, $J^T_x(q_y)$ the transverse current
operator (equivalently one can choose the transverse current operator $J^T_x(q_z)$), and $|n\rangle$ is a set of many-body eigenstates with
energy $E_n$. 

Differently from the $y$ and $z$ components, the $x$ component of the current operator is affected by SOC according to
\begin{equation}
J^T_x(q_y) = \sum_k (p_{x,k} - \hbar k_R \sigma_{z,k}) \me^{\mi q_y y_k}
\label{eq:trans_curr}
\end{equation}
(here the sum runs over the constituent atoms of the gas) so that the normal density (and hence the superfluid density) is an anisotropic
diagonal tensor, whose transverse components $\rho_s^y = \rho_s^z$ match the total mass density $\rho$ at zero temperature, while the
longitudinal component $\rho_s^x$ is smaller than $\rho$ even at $T=0$.

\begin{figure}
\centering
\includegraphics{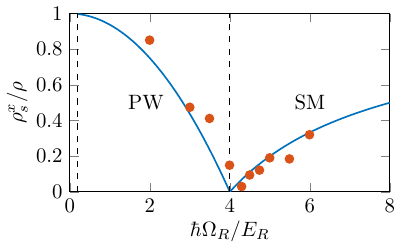}
\caption{Superfluid density along $x$ of a spin-orbit-coupled BEC in the PW and SM phases as a function of the Raman coupling. The solid curve represents
the theoretical prediction~\eqref{eq:rho_s_kappa}. The red circles represent experimental values deduced from Eq.~\eqref{eq:rho_s_kappa}, using the
measured sound velocities taken from~\cite{Ji2015} and the value of the compressibility calculated in~\cite{Martone2012}. The vertical dashed lines
pinpoint the transitions between the various phases. Adapted from \cite{Zhang2016}.}
\label{fig:superf}
\end{figure}

Since the transverse current operator does not excite the gapless phonon mode, which is of longitudinal nature, the only contribution to
Eq.~\eqref{eq:norm_dens} comes from the gapped branch, and in the evaluation of such a contribution one can safely take the $q \to 0$ limit of the
current operator, which reduces to the simple form $J^T_x(q=0)= \sum_k(p_{x,k} - \hbar k_R \sigma_{z,k})$. Since the canonical momentum
$\sum_k p_{x,k}$ commutes with the Hamiltonian and cannot excite states with $E_n \neq 0$, one finds that only the spin component of the current
operator plays a role in the calculation of the normal density. One eventually obtains the nontrivial result~\cite{Zhang2016}
\begin{equation}
\frac{\rho_s^x}{\rho}= \frac{1}{1 + 2 E_R \chi_M}
\label{eq:rho_s_x}
\end{equation}
for the superfluid density, whose behavior is fixed by that of the magnetic susceptibility $\chi_M$, already discussed in Sec.~\ref{sec:gs}. We recall,
in particular, that $\chi_M$ exhibits a divergent behavior as one approaches the transition between the PW and the SM phase. Furthermore,
if $g_{ss} = 0$, the magnetic susceptibility is fixed by the effective mass according to Eq.~\eqref{eq:eff_mass} and the expression for the superfluid
density reduces to $\rho_s^x / \rho = m/m^*$. Figure~\ref{fig:superf} reports the value of the superfluid density calculated as a function of the Raman
coupling and revealing the dramatic suppression of $\rho_s^x$ in an important interval of the physical parameter $\Omega_R$. These results can be extended
to the supersolid stripe phase~\cite{Martone2021} and have been confirmed by calculations based on the use of the phase twist approach~\cite{Chen2018}
and Monte Carlo simulations~\cite{Sanchez2020}.

The superfluid density has an important effect on the propagation of sound, as naturally expected from general hydrodynamic arguments applied to
superfluid systems. By employing sum-rule arguments, one can actually show that in the case of spin-orbit-coupled BECs, the superfluid density is related
to the compressibility $\kappa$ and the sound velocities $c_{x,\pm}$ introduced in Sec.~\ref{sec:hd} as
\begin{equation}
\frac{\rho_s^x}{\rho} = m c_{x,+} c_{x,-} \kappa \, .
\label{eq:rho_s_kappa}
\end{equation}
This result, combined with the standard relation $m c^2_\perp = \kappa^{-1}$ holding along the transverse direction, suggests that the superfluid density
could be actually measured by studying the ratio  
\begin{equation}
\frac{\rho_s^x}{\rho} = \frac{c_{x,+}c_{x,-}}{c_\perp^2}
\label{eq:rho_s_c}
\end{equation}
between the sound velocities along and perpendicular to the $x$ direction~\cite{Martone2021}. The suppression of the superfluid density along the $x$
direction, predicted in~\cite{Zhang2016}, was confirmed by the measurements of the sound velocities $c_{x,+}$ and $c_{x,-}$ reported in~\cite{Ji2015}.
In addition, the ratio~\eqref{eq:rho_s_c} between the longitudinal and transverse sound velocities was recently successfully measured to determine
the superfluid fraction of a single-component BEC gas in the presence of a periodically modulated potential~\cite{Tao2023,Chauveau2023}.

A similar investigation can be carried out for the moment of inertia, whose reduction with respect to the rigid value is considered an important indicator
of superfluidity. The calculation of the moment of inertia is based on the determination of the response of the system to a static rotational constraint of
the form $- \omega_{\mathrm{rot}} L_z$, where
\begin{equation}
L_z = - \mi \hbar (x \nabla_y - y \nabla_x) + \hbar k_R y \sigma_z
\label{eq:ang_mom}
\end{equation}
is the angular momentum operator, which exhibits an explicit spin dependence due to the modification of the momentum operator induced by SOC (see
Sec.~\ref{sec:symm}). The calculation of the moment of inertia $\Theta = \langle L_z \rangle / \omega_{\mathrm{rot}}$ can be handled in analytic form
by employing the hydrodynamic description developed in the previous section for the SM phase within the $g_{ss} = 0$ approximation. When applied to an
isotropically trapped configuration in the $xy$ plane, the rotational constraint affects only the equation~\eqref{eq:hd_novel} fixing the relationship
between the gradient of the phase and the spin density, which takes the novel form:
\begin{equation}
- \frac{\hbar k_R}{m} \nabla_x \phi + \frac{\Omega_R}{2} \frac{s_z}{n} - \omega_{\mathrm{rot}} k_R y = 0 \, .
\label{eq:rot_novel}
\end{equation}
The hydrodynamic equations are easily solved in this case~\cite{Stringari2017}, and one finds that in the SM phase, the velocity field $\vec{v} =
\vec{j} / n$ takes the peculiar rigid-like expression $\vec{v} = (\vec{\omega}_{\mathrm{rot}} \times \vec{r}) \Omega_{\mathrm{cr2}} /
(2\Omega_R - \Omega_{\mathrm{cr2}})$ yielding the result (a similar expression holds in the PW phase~\cite{Stringari2017})
\begin{equation}
\frac{\Theta}{\Theta_{\mathrm{rig}}} = \frac{\Omega_{\mathrm{cr2}}}{2 \Omega_R - \Omega_{\mathrm{cr2}}}
\label{eq:mom_in}
\end{equation}
for the moment of inertia. Remarkably, at the transition $\Omega_R = \Omega_{\mathrm{cr2}}$ between the two phases, one finds $\vec{v} =
\vec{\omega}_{\mathrm{rot}} \times \vec{r}$ and $\Theta = \Theta_{\mathrm{rig}}$, i.e., the system rotates like a rigid body, violating the usual
irrotationality constraint of single-component superfluids.

\section{Stripe phase and supersolidity}
\label{sec:stripe}
The phenomenon of supersolidity occurs when both global-phase and translation invariance are spontaneously broken in a quantum system, which
consequently exhibits superfluidity and spatial order at the same time~\cite{Thouless1969,Andreev1969,Leggett1970,Kirzhnits1971}. After several
unsuccessful attempts to detect this behavior in solid helium~\cite{Balibar2010review,Boninsegni2012review}, supersolid configurations have been
observed in ultracold Bose gases inside two optical resonators~\cite{Leonard2017}, with spin-orbit coupling~\cite{Li2017,Putra2020}, and with
dipole-dipole interactions~\cite{Tanzi2019a,Boettcher2019,Chomaz2019}. In particular, the stripe phase of spin-orbit-coupled BECs has emerged
as a promising candidate for exploring supersolid effects. As discussed in Sec.~\ref{sec:gs}, this phase is predicted to occur at small Raman
coupling, below the critical value~\eqref{eq:omega_cr1}. It is the consequence of interaction effects, and in particular results from the competition
between the density-density and spin-spin interaction terms entering the energy~\eqref{eq:gp_en} in the presence of the spin-orbit single-particle
Hamiltonian~\eqref{eq:h_SO}. The stripe phase is also called the phase-mixed configuration as the condensate occupies an equal-weighted combination
of the two spinor states in the Ansatz~\eqref{eq:ansatz}. These states thus coexist in all space, yielding a zero spin density $s_z$ everywhere and
minimizing the spin-spin interaction term of Eq.~\eqref{eq:gp_en} (if $g_{ss} > 0$). Since the two spinor states carry opposite wave vectors
$\pm k_1 \hat{\vec{e}}_x$, by interfering they give rise to fringes in the density profile with wavelength $\pi / k_1$. The contrast of the fringes,
defined as $\mathcal{C} = (n_{\mathrm{max}} - n_{\mathrm{min}}) / (n_{\mathrm{max}} + n_{\mathrm{min}})$ with $n_{\mathrm{min}}$ ($n_{\mathrm{max}}$)
the minimum (maximum) value of the density, increases linearly with the Raman coupling, being given by the equation~\cite{Li2012a}
\begin{equation}
\mathcal{C} = \frac{\hbar \Omega_R}{4 E_R + g_{dd} \bar{n}} \, .
\label{eq:contrast}
\end{equation}
The energy cost of fringes becomes then higher and higher at increasing $\Omega_R$, being mainly associated with the density-density interaction term
of Eq.~\eqref{eq:gp_en}, and eventually the system undergoes a first-order phase transition to the PW phase, as already pointed out in Sec.~\ref{sec:gs}.
This transition was already identified in the first pioneering experiment of the NIST group~\cite{Lin2011}, where the system was initially prepared
in an unpolarized mixture of spin-up and spin-down atoms at $\Omega_R = 0$, and then the Raman coupling was adiabatically ramped up. Above the
transition point (whose observed value $\hbar\Omega_{\mathrm{cr1}} \simeq 0.19 \, \! E_R$ agrees very well with the theoretical
prediction~\eqref{eq:omega_cr1}) the populations of the two spinor states of the wave function~\eqref{eq:ansatz} separated in different spatial domains,
with the system as a whole staying unpolarized~\cite{Lin2011,Putra2020} (this is why the PW phase is sometimes referred to as the phase-separated
configuration). More recently, it was shown that in this regime, density modulations can still be detected in the domain wall, where there is a residual
overlap of the two components~\cite{Putra2020}.

A major problem of experiments with $^{87}$Rb atom is represented by the tiny difference between the intra- and interspecies coupling constants,
entailing the smallness of the ratio $g_{ss} / g_{dd} \sim 10^{-3}$ and hence of the critical value $\Omega_{\mathrm{cr1}}$ (see above).
This strongly reduces the available range of Raman couplings for observing the stripe phase, which cannot exceed $\Omega_{\mathrm{cr1}}$, thus inhibiting
the realization of fringes with large contrast~\cite{Martone2014}. In addition, in these conditions the density fringes can be easily destroyed by
fluctuations of external magnetic fields~\cite{Martone2015}. The first detection of the crystal features of the stripe phase was actually achieved by
first implementing SOC in pseudospin configurations generated by a superlattice~\cite{Li2016}, and then using Bragg scattering~\cite{Li2017}. The Bragg
signal was later measured also in a SOC mixture of $^{87}$Rb atoms, by amplifying the contrast of fringes via a sudden increase of the Raman coupling,
see Ref.~\cite{Putra2020} (where the presence of long-range spatial coherence in spin-orbit-coupled BECs was also established). In the above experiments,
the direct observation of fringes in the density profile is inhibited by the narrow separation between stripes, whose value is mainly determined by the
magnitude $2 \hbar k_R$ of the momentum transfer of the laser fields. Very recently, the experimental observation of density fringes has been reported
in a spin-orbit-coupled mixture of Potassium atoms, employing a matter-wave lensing technique, which amplifies the space separation between the
stripes~\cite{Tarruell2023}.

\begin{figure}
\centering
\includegraphics{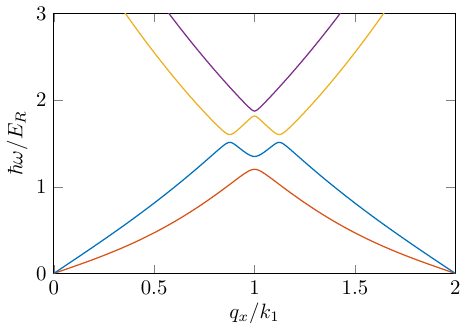}
\caption{Lowest four bands of the Bogoliubov dispersion of a spin-orbit-coupled BEC in the stripe phase as functions of the quasimomentum. At small $q_x$,
the two lowest bands feature density (blue curve) and spin (red curve) phonon modes. Notice also the vanishing of their frequencies at the Brillouin
point $q_x = 2 k_1$. Adapted from~\cite{Martone2021} under the terms of the Creative Commons Attribution 4.0 International License (see
\texttt{http://creativecommons.org/licenses/by/4.0/}).}
\label{fig:stripe}
\end{figure}

\begin{figure}
\centering
\includegraphics{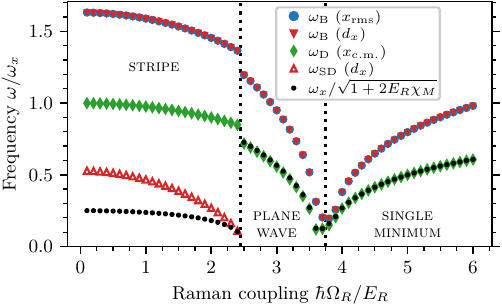}
\caption{Frequency of the breathing mode ($\omega_{\mathrm{B}}$), the spin-dipole mode ($\omega_{\mathrm{SD}}$), and the center-of-mass (dipole) mode
($\omega_{\mathrm{D}}$) of a trapped spin-orbit-coupled BEC as a function of the Raman coupling. The value of each frequency is deduced by looking at
the oscillations of the center-of-mass coordinate, $x_{\mathrm{c.m.}} = \langle x \rangle$, the relative displacement of the two spin components,
$d_x = \langle x \sigma_z \rangle$, and the root-mean-square radius of the atomic cloud in the $x$ direction, $x_{\mathrm{rms}} =
\sqrt{\langle x^2 \rangle}$. The upper bound~\eqref{eq:dipole_chi} for $\omega_{\mathrm{D}}$ is also shown. Following the sudden removal of a small
static perturbation proportional to the operator $x^2$, in the PW and SM phases, $d_x$ and $x_{\mathrm{rms}}$ oscillate at a single frequency equal to
$\omega_{\mathrm{B}}$, revealing the full hybridization of the breathing and spin-dipole modes. Conversely, in the stripe phase, the two modes remain
distinct, although they acquire a hybrid density and spin character, which results in a beating in the signal $d_x$. Adapted from~\cite{Geier2021}.}
\label{fig:trap_modes}
\end{figure}

At the level of dynamics, the stripe phase of spin-orbit-coupled Bose gases exhibits novel properties with respect to supersolids in single-component
systems such as dipolar BECs~\cite{Natale2019,Tanzi2019b,Guo2019,Petter2021}. In particular, in spin-orbit-coupled BECs, the Goldstone mode associated with
the spontaneous breaking of translational invariance has a typical spin nature, which at the same time incapsulates its crystal character. This can be seen
by computing the frequencies $\omega$ of the Bogoliubov modes and the corresponding amplitudes $U(\vec{r})$ and $V(\vec{r})$ (see Sec.~\ref{sec:hd}),
which in infinite setups can be taken as Bloch waves with well-defined excitation quasimomentum $\hbar\vec{q}$~\cite{Li2013,Martone2021}. The excitation
spectrum has a band structure whose two lowest bands are gapless and take the form, at small wave vectors, of density and spin phonons, see
Fig.~\ref{fig:stripe}. It has been proven~\cite{Martone2021} that, while the zero-frequency density mode only shifts the value of the condensate global
phase, the Goldstone mode in the spin band is associated with a rigid translational motion of stripes; this mode can be excited by a sudden change of the
magnetic detuning $\delta_R$ (see Eq.~\eqref{eq:h_SO}), both in infinite and trapped systems~\cite{Geier2021,Geier2023}. On the other hand, at finite
$\vec{q}$, the spin phonon has been recently shown~\cite{Geier2023} to be characterized by a typical oscillation of the relative distance between stripes
(if $\vec{q}$ lies along $x$) or of their orientation (for $\vec{q}$ perpendicular to $x$), thus explicitly revealing its crystal nature. In the presence of
harmonic trapping, a similar crystal oscillation of stripes characterizes the spin-dipole mode, whose frequency becomes smaller and smaller as one
approaches the transition to the PW phase (see Fig.~\ref{fig:trap_modes}). This discussion shows that the spin degree of freedom featured by BEC mixtures
with SOC can open unprecedented experimental perspectives for the excitation and monitoring of the novel Goldstone modes exhibited by the supersolid state
of matter.

We finally point out that the frequencies of the two lowest bands of the spectrum of Fig.~\ref{fig:stripe} also vanish at $q_x = 2 k_1$, i.e., the edge of
the first Brillouin zone. A consequence of this feature is that the motion of an external body in the direction perpendicular to the fringes can never be
frictionless, regardless of how small the velocity is~\cite{Martone2018}. Notice, however, that the superfluid fraction in the stripe phase is
finite~\cite{Chen2018,Martone2021}, although its value is always smaller than one because of the simultaneous breaking of both translation and Galilean
symmetry (see Secs.~\ref{sec:symm} and~\ref{sec:sup_rot}).

\section{Conclusion}
\label{sec:concl}
In these lecture notes, we have discussed some of the most prominent features of spin-orbit-coupled atomic Bose gases. By exploiting the coupling between
the atoms and a properly designed configuration of Raman lasers, one can engineer a single-particle dispersion possessing a double-minimum structure. In
the presence of interactions, this gives rise to a complex phase diagram, including uniform phases where the atoms condense in a state with finite or zero
momentum, and a stripe phase with a supersolid behavior. These phases exhibit interesting dynamic features, such as the strong suppression of the collective
mode frequencies (both in infinite and trapped configurations) close to the second-order transition between the plane-wave and the single-minimum phases,
a roton minimum in the plane-wave phase, and a band structure with two gapless branches in the stripe phase. In addition, the superfluidity and rotational
properties of the gas are also strikingly modified in the presence of spin-orbit coupling. The study of these properties, as well as of the phonon modes of
the stripe phase, revealing of its dynamic crystal character, is well under reach of current experiments and represents a potential future direction of
research in this field.

\acknowledgments
Many useful discussions and collaborations with K. T. Geier, P. Hauke, W. Ketterle, and Y. Li are acknowledged.
This work was supported by the Italian Ministry of University and Research (MUR) through the PRIN project INPhoPOL
(grant 2017P9FJBS) and the PNRR MUR project PE0000023 - NQSTI.


\begin{thebibliography}{0}

\bibitem{Dalibard2011review}
\BY{Dalibard J., Gerbier F., Juzeli\={u}nas G. \atque \"{O}hberg P.}
\IN{Rev. Mod. Phys.}{83}{2011}{1523}.

\bibitem{Goldman2014review}
\BY{Goldman N., Juzeli\={u}nas G., \"{O}hberg P. \atque Spielman I. B.}
\IN{Rep. Prog. Phys.}{77}{2014}{126401}.

\bibitem{Lin2009}
\BY{Lin Y.-J., Compton R. L., Jimenez-Garcia K., Porto J. V. \atque Spielman I. B.}
\IN{Nature (London)}{462}{2009}{628}.

\bibitem{Lin2011}
\BY{Lin Y.-J., Jimenez-Garcia K. \atque Spielman I. B.}
\IN{Nature (London)}{471}{2011}{83}.

\bibitem{Galitski2013review}
\BY{Galitski V. \atque Spielman I. B.}
\IN{Nature (London)}{494}{2013}{49}.

\bibitem{Zhou2013review}
\BY{Zhou X., Li Y., Cai Z. \atque Wu C.}
\IN{J. Phys. B}{46}{2013}{134001}.

\bibitem{Zhai2015review}
\BY{H. Zhai}
\IN{Rep. Prog. Phys.}{78}{2015}{026001}.

\bibitem{Li2015review}
\BY{Li Y., Martone G. I. \atque Stringari S.}
\TITLE{Spin-Orbit-Coupled Bose-Einstein Condensates},
in \TITLE{Annual Review of Cold Atoms and Molecules}
edited by \NAME{K. W. Madison, K. Bongs, L. D. Carr, A. M. Rey, H. Zhai}
(World Scientific, Singapore), Vol. 3, 2015, pp.~201-250.

\bibitem{Zhang2016review}
\BY{Zhang Y., Mossman M. E., Busch T., Engels P. \atque Zhang C.}
\IN{Front. Phys.}{11}{2016}{118103}.

\bibitem{Recati2022review}
\BY{Recati A. \atque Stringari S.}
\TITLE{Coherently Coupled Mixtures of Ultracold Atomic Gases},
in \TITLE{Annual Review Of Condensed Matter Physics}
edited by \NAME{A. P. Mackenzie, M. C. Marchetti}
(Annual Reviews, San Mateo), Vol. 13, 2022, pp.~407-432.

\bibitem{Martone2023review}
\BY{Martone G. I.}
\IN{EPL}{143}{2023}{25001}.

\bibitem{Li2017}
\BY{Li J., Lee J., Huang W., Burchesky S., Shteynas B., Top F. \c{C}., Jamison A. O. \atque Ketterle W.}
\IN{Nature (London)}{543}{2017}{91}.

\bibitem{Martone2012}
\BY{Martone G. I., Li Y., Pitaevskii L. P. \atque Stringari S.}
\IN{Phys. Rev. A}{86}{2012}{063621}.

\bibitem{Bychkov1984}
\BY{Bychkov Y. A. \atque Rashba E. I.}
\IN{J. Phys. C}{17}{1984}{6039}.

\bibitem{Dresselhaus1955}
\BY{Dresselhaus G.} 
\IN{Phys. Rev.}{100}{1955}{580}.

\bibitem{Zhu2012}
\BY{Zhu Q., Zhang C. \atque Wu B.}
\IN{EPL}{100}{2012}{50003}.

\bibitem{Zheng2013}
\BY{Zheng W., Yu Z.-Q., Cui X. \atque Zhai H.}
\IN{J. Phys. B}{46}{2013}{134007}

\bibitem{Ozawa2013}
\BY{Ozawa T., Pitaevskii L. P. \atque Stringari S.}
\IN{Phys. Rev. A}{87}{2013}{063610}.

\bibitem{Pitaevskii_Stringari_book}
\BY{Pitaevskii L. P. \atque Stringari S.}
\TITLE{Bose-Einstein Condensation and Superfluidity}
(Oxford University Press, Oxford, 2016).

\bibitem{Ho2011}
\BY{T.-L. Ho \atque S. Zhang}
\IN{Phys. Rev. Lett.}{107}{2011}{150403}.

\bibitem{Li2012a}
\BY{Li Y., Pitaevskii L. P. \atque Stringari S.}
\IN{Phys. Rev. Lett.}{108}{2012}{225301}.

\bibitem{Wang2010}
\BY{Wang C., Gao C., Jian C.-M. \atque Zhai H.}
\IN{Phys. Rev. Lett.}{105}{2010}{160403}.

\bibitem{Wu2011}
\BY{Wu C.-J., Mondragon-Shem I. \atque Zhou X.-F.}
\IN{Chin. Phys. Lett.}{28}{2011}{097102}..

\bibitem{Li2013}
\BY{Li Y., Martone G. I., Pitaevskii L. P. \atque Stringari S.}
\IN{Phys. Rev. Lett.}{110}{2013}{235302}.

\bibitem{Martone2021}
\BY{Martone G. I. \atque Stringari S.}
\IN{SciPost Phys.}{11}{2021}{092}.

\bibitem{Hamner2014}
\BY{Hamner C., Qu C., Zhang Y., Chang J., Gong M., Zhang C. \atque Engels P.}
\IN{Nat. Commun.}{5}{2014}{4023}.

\bibitem{Chen2018}
\BY{Chen X.-L., Wang J., Li Y., Liu X.-J. \atque Hu H.}
\IN{Phys. Rev. A}{98}{2018}{013614}.

\bibitem{Sanchez2020}
\BY{S\'{a}nchez-Baena J., Boronat J. \atque Mazzanti F.}
\IN{Phys. Rev. A}{101}{2020}{043602}.

\bibitem{Yu2014}
\BY{Yu Z.-Q.}
\IN{Phys. Rev. A}{90}{2014}{053608}.

\bibitem{Ji2014}
\BY{Ji S.-C., Zhang J.-Y., Zhang L., Du Z.-D., Zheng W., Deng Y.-J., Zhai H., Chen S. \atque Pan J.-W.}
\IN{Nat. Phys.}{10}{2014}{314}.

\bibitem{Li2012b}
\BY{Li Y., Martone G. I. \atque Stringari S.}
\IN{EPL}{99}{2012}{56008}.

\bibitem{Zhang2012}
\BY{Zhang J.-Y., Ji S.-C., Chen Z., Zhang L., Du Z.-D., Yan B., Pan G.-S., Zhao B., Deng Y.-J., Zhai H., Chen S. \atque Pan J.-W.}
\IN{Phys. Rev. Lett.}{109}{2012}{115301}.

\bibitem{Parker2013}
\BY{Parker C. V., Ha L.-C. \atque Chin C.}
\IN{Nat. Phys.}{9}{2013}{769}.

\bibitem{Ha2015}
\BY{Ha L.-C., Clark L. W., Parker C. V., Anderson B. M. \atque Chin C.}
\IN{Phys. Rev. Lett.}{114}{2015}{055301}.

\bibitem{Zheng2012}
\BY{Zheng W. \atque Li Z.}
\IN{Phys. Rev. A}{85}{2012}{053607}.

\bibitem{Ji2015}
\BY{Ji S.-C., Zhang L., Xu X.-T., Wu Z., Deng Y., Chen S. \atque Pan J.-W.}
\IN{Phys. Rev. Lett.}{114}{2015}{105301}.

\bibitem{Khamehchi2014}
\BY{Khamehchi M. A., Zhang Y., Hamner C., Busch T. \atque Engels P.}
\IN{Phys. Rev. A}{90}{2014}{063624}.

\bibitem{Chen2017}
\BY{Chen L., Pu H., Yu Z.-Q. \atque Zhang Y.}
\IN{Phys. Rev. A}{95}{2017}{033616}.

\bibitem{Geier2021}
\BY{Geier K. T., Martone G. I., Hauke P. \atque Stringari S.}
\IN{Phys. Rev. Lett.}{127}{2021}{115301}.

\bibitem{Zhang2016}
\BY{Zhang Y.-C., Yu Z.-Q., Ng T. K., Zhang S., Pitaevskii L. P. \atque Stringari S.}
\IN{Phys. Rev. A}{94}{2016}{033635}.

\bibitem{Tao2023}
\BY{Tao J., Zhao M. \atque Spielman I. B.}
\IN{Phys. Rev. Lett.}{131}{2023}{163401}.

\bibitem{Chauveau2023}
\BY{Chauveau G., Maury C., Rabec F., Heintze C., Brochier G., Nascimbene S.,
Dalibard J., Beugnon J., Roccuzzo S. M. \atque Stringari S.}
\IN{Phys. Rev. Lett.}{130}{2023}{226003}.

\bibitem{Stringari2017}
\BY{Stringari S.}
\IN{Phys. Rev. Lett.}{118}{2017}{145302}.

\bibitem{Thouless1969}
\BY{Thouless D. J.}
\IN{Ann. Phys. (N.Y.)}{52}{1969}{403}.

\bibitem{Andreev1969}
\BY{Andreev A. F. \atque Lifshitz I. M.}
\IN{Sov. Phys. JETP}{29}{1969}{1107}.

\bibitem{Leggett1970}
\BY{Leggett A. J.}
\IN{Phys. Rev. Lett.}{25}{1970}{1543}.

\bibitem{Kirzhnits1971}
\BY{Kirzhnits D. A. \atque Nepomnyashchii Y. A.}
\IN{Sov. Phys. JETP}{32}{1971}{1191}.

\bibitem{Balibar2010review}
\BY{Balibar S.}
\IN{Nature (London)}{464}{2010}{176}.

\bibitem{Boninsegni2012review}
\BY{Boninsegni M. \atque Prokof'ev N. V.}
\IN{Rev. Mod. Phys.}{84}{2012}{759}.

\bibitem{Leonard2017}
\BY{L\'{e}onard J., Morales A., Zupancic P., Esslinger T. \atque Donner T.}
\IN{Nature (London)}{543}{2017}{87}.

\bibitem{Putra2020}
\BY{Putra A., Salces-C\'{a}rcoba F., Yue Y., Sugawa S. \atque Spielman I. B.}
\IN{Phys. Rev. Lett.}{124}{2020}{053605}.

\bibitem{Tanzi2019a}
\BY{Tanzi L., Lucioni E., Fam\`{a} F., Catani J., Fioretti A., Gabbanini C., Bisset R. N.,
Santos L. \atque Modugno G.}
\IN{Phys. Rev. Lett.}{122}{2019}{130405}.

\bibitem{Boettcher2019}
\BY{B\"{o}ttcher F., Schmidt J.-N., Wenzel M., Hertkorn J., Guo M., Langen T. \atque Pfau T.}
\IN{Phys. Rev. X}{9}{2019}{011051}.

\bibitem{Chomaz2019}
\BY{Chomaz L., Petter D., Ilzh\"{o}fer P., Natale G., Trautmann A., Politi C.,
Durastante G., van Bijnen R. M. W., Patscheider A., Sohmen M., Mark M. J. \atque Ferlaino F.}
\IN{Phys. Rev. X}{9}{2019}{021012}.

\bibitem{Martone2014}
\BY{Martone G. I., Li Y. \atque Stringari S.}
\IN{Phys. Rev. A}{90}{2014}{041604}.

\bibitem{Martone2015}
\BY{Martone G. I.}
\IN{Eur. Phys. J. Special Topics}{224}{2015}{553}.

\bibitem{Li2016}
\BY{Li J., Huang W., Shteynas B., Burchesky S., Top F. \c{C}., Su E., Lee J., Jamison A. O. \atque Ketterle W.}
\IN{Phys. Rev. Lett.}{117}{2016}{185301}.

\bibitem{Tarruell2023}
\BY{Tarruell L.}
private communication.

\bibitem{Natale2019}
\BY{G. Natale, R. M. W. van Bijnen, A. Patscheider, D. Petter, M. J. Mark,
L. Chomaz \atque F. Ferlaino}
\IN{Phys. Rev. Lett.}{123}{2019}{050402}.

\bibitem{Tanzi2019b}
\BY{Tanzi L., Roccuzzo S. M., Lucioni E., Fam\`{a} F., Fioretti A.,
Gabbanini C., Modugno G., Recati A. \atque Stringari S.}
\IN{Nature (London)}{574}{2019}{382}.

\bibitem{Guo2019}
\BY{Guo M., B\"{o}ttcher F., Hertkorn J., Schmidt J.-N., Wenzel M.,
B\"{u}chler H. P., Langen T. \atque Pfau T.}
\IN{Nature (London)}{574}{2019}{386}.

\bibitem{Petter2021}
\BY{Petter D., Patscheider A., Natale G., Mark M. J., Baranov M. A., van Bijnen R.,
Roccuzzo S. M., Recati A., Blakie B., Baillie D., Chomaz L. \atque Ferlaino F.}
\IN{Phys. Rev. A}{104}{2021}{L011302}.

\bibitem{Geier2023}
\BY{Geier K. T., Martone G. I., Hauke P., Ketterle W. \atque Stringari S.}
\IN{Phys. Rev. Lett.}{130}{2023}{156001}.

\bibitem{Martone2018}
\BY{Martone G. I. \atque Shlyapnikov G. V.}
\IN{J. Exp. Theor. Phys.}{127}{2018}{865}.

\end{thebibliography}
\end{document}